\documentclass[]{interact}

\usepackage{epstopdf}
\usepackage[caption=false]{subfig}

\usepackage{natbib}
\bibpunct[, ]{(}{)}{;}{a}{}{,}
\renewcommand\bibfont{\fontsize{10}{12}\selectfont}

\theoremstyle{plain}

\theoremstyle{definition}

\theoremstyle{remark}

\begin{document}


\title{Phase and Intensity Control of Dissipative Kerr Cavity Solitons}

\author{
\name{Miro Erkintalo\textsuperscript{a,b}\thanks{CONTACT M. Erkintalo. Email: m.erkintalo@auckland.ac.nz}, Stuart~G. Murdoch\textsuperscript{a,b} and St\'ephane Coen\textsuperscript{a,b}}
\affil{\textsuperscript{a}Department of Physics, The University of Auckland, Auckland 1010, New Zealand; \textsuperscript{b}The Dodd-Walls Centre for Photonic and Quantum Technologies, New Zealand}
}

\maketitle

\begin{abstract}
Dissipative Kerr cavity solitons are pulses of light that can persist in coherently driven nonlinear optical resonators. They have attracted significant attention over the past decade due to their rich nonlinear dynamics and key role in the generation of coherent microresonator optical frequency combs. Whilst the vast majority of implementations have relied on \emph{homogeneous} continuous wave driving, the soliton's ``plasticity'' combined with \emph{inhomogeneous} driving offers attractive advantages for a host of applications. Here we review recent studies into the dynamics and applications of Kerr cavity solitons in the presence of inhomogeneous driving fields. In particular, we summarise the salient theoretical developments that allow for the analysis of CS motion in the presence of pump phase or amplitude inhomogeneities, and survey recent experiments that use pulsed driving to realise energy efficient and flexible microresonator optical frequency combs.
\end{abstract}


\section{Introduction}
An externally-driven, passive optical resonator with Kerr nonlinearity can support persisting pulses of light known as temporal cavity solitons (CSs; also known as dissipative Kerr solitons)~\citep{wabnitz_suppression_1993,leo_temporal_2010, herr_temporal_2014,kippenberg_dissipative_2018}. Such CSs are (typically ultrashort) pulses of light, sitting atop a low-level homogeneous background, and they can recirculate around the resonator indefinitely. They are attractors of the underlying nonlinear dynamical system, and their persistence is underpinned by a double-balance~\citep{akhmediev_dissipative_2008}: all the energy they lose is replenished via parametric gain from the background on top of which they sit (and that is externally replenished by the driving field) and their dispersive spreading is cancelled by the self-focussing Kerr nonlinearity.

Analogous to spatial localized structures that can manifest themselves in diffractive resonators~\citep{barland_cavity_2002,ackemann_chapter_2009}, temporal CSs were first observed in a macroscopic optical fibre ring resonator~\citep{leo_temporal_2010} and soon thereafter in a monolithic microresonator~\citep{herr_temporal_2014}.  The former (macroscopic) systems have proven invaluable for the controlled exploration of the characteristics and rich nonlinear dynamics of CSs~\citep{leo_dynamics_2013,anderson_observations_2016,anderson_coexistence_2017,wang_stimulated_2018,nielsen_coexistence_2019,xue_super-efficient_2019}, but it is the latter (microscopic) systems that have unleashed the solitons' application potential~\citep{kippenberg_dissipative_2018, pasquazi_micro-combs:_2018}. In particular, it is now well-understood that temporal CSs represent the time-domain counterparts of the coherent optical frequency combs that can be generated in Kerr nonlinear microresonators~\citep{herr_temporal_2014, brasch_photonic_2016, yi_soliton_2015, joshi_thermally_2016, jang_synchronization_2018}. Such soliton microresonator frequency combs represent one of the most important optical technologies to emerge in the past decade, with demonstrated applications including high-speed telecommunications~\citep{marin-palomo_microresonator-based_2017}, optical distance measurements~\citep{trocha_ultrafast_2018,suh_soliton_2018, riemensberger_massively_2020}, high-resolution spectroscopy~\citep{suh_microresonator_2016, dutt_-chip_2018, lucas_spatial_2018, spencer_optical-frequency_2018}, and detection of extrasolar planets~\citep{suh_searching_2019,obrzud_microphotonic_2019}.

\begin{figure}[!t]
 \centering
  \includegraphics[width = \textwidth, clip=true]{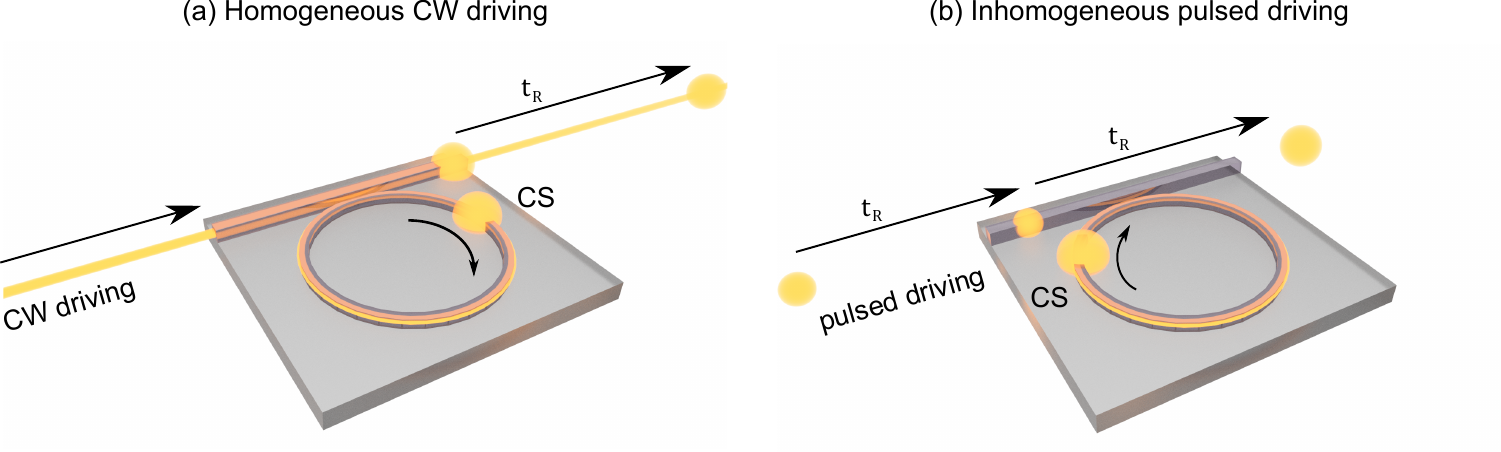}
 \caption{Schematic illustration of CS-supporting nonlinear resonator systems with (a) homogeneous CW and (b) inhomogeneous pulsed driving. In (a), a CW driving field produces a periodic train of ultrashort CS pulses that repeat at the cavity round trip time $t_\mathrm{R}$. In (b), a pulsed driving field with periodicity $t_\mathrm{R}$ synchronously drives the resonator and produces intense ultrashort CS pulses with the same periodicity.}
 \label{fig1}
\end{figure}

Most CS systems use purely homogeneous, continuous wave (CW) laser radiation to drive the resonator [see Fig.~\ref{fig1}(a)]. Whilst simple to implement, CW driving suffers from certain shortcomings~\citep{obrzud_temporal_2017, xu_harmonic_2020}. First, CW driving fills the entire cavity with a homogeneous ``holding'' field, yet CSs only extract energy from the portion of the field with which they overlap. As the duration of that portion typically corresponds to a very small fraction of the cavity round trip time, a considerable amount of energy is effectively wasted, resulting in poor pump-to-soliton conversion efficiency~\citep{bao_nonlinear_2014, xue_microresonator_2017}. Second, with pure CW driving, it can be challenging (albeit not impossible) to control the specific soliton configuration that circulates in the resonator~\citep{guo_universal_2017}, making it in particular difficult to reach the single-soliton state that is arguably of most applied interest. Third, CW driving does not provide an intrinsic means to lock the soliton repetition rate to an external radiofrequency (RF) reference, which is needed for many applications.

The issues listed above have stimulated increasing interest in driving scenarios that go beyond the simplest paradigm of pure CW driving [see Fig.~\ref{fig1}(b)]. In particular, it is now known that the introduction of phase or amplitude inhomogeneities on the cavity driving field allows CSs to be robustly trapped at specific temporal positions~\citep{jang_temporal_2015, obrzud_temporal_2017,  hendry_spontaneous_2018, cole_kerr-microresonator_2018}, thus permitting control over the soliton configuration as well as their repetition rate. Moreover, the use of a pulsed driving source that is synchronised with the cavity round trip time has been shown to provide an attractive route to enhancing the pump-to-soliton conversion efficiency~\citep{obrzud_temporal_2017}, and also to enable CS comb generation in resonators that would not be suitable under CW driving~\citep{lilienfein_temporal_2019, xu_harmonic_2020}.

Here we review recent studies of CS dynamics in the presence of driving field inhomogeneities. In particular, we recount the key steps of a comprehensive theoretical description by Maggipinto et al. to elucidate the motion (or lack thereof) of CSs in the presence of phase or amplitude inhomogeneities~\citep{maggipinto_cavity_2000}; describe how the appropriate CS drift velocities can be calculated in practice~\citep{hendry_spontaneous_2018}; and survey recent experimental results.

\section{Theory of CS motion}
\label{theorysec}
We begin by recalling the theory of CS motion under inhomogeneous driving fields. Whilst we focus on \emph{temporal} CSs in \emph{dispersive} resonators, our analysis follows closely the work of Maggipinto et al. that was originally presented in the context of \emph{spatial} CSs in \emph{diffractive} semiconductor microcavities~\citep{maggipinto_cavity_2000}. We must also emphasise that, whilst our focus is on \emph{external} inhomogeneities arising from the cavity driving fields, the general theory presented below can also be extended to \emph{internal} perturbations~\citep{maggipinto_cavity_2000} that underpin, e.g., the formation of bound soliton states~\citep{wang_universal_2017} and soliton crystals~\citep{cole_kerr-microresonator_2018,karpov_dynamics_2019}.

It is well know that the dynamics of temporal CSs in dispersive Kerr resonators can be described by the mean-field Lugiato-Lefever equation~\citep{lugiato_spatial_1987}, which was first derived in the context of dispersive resonators by Haelterman et al.~\citep{haelterman_dissipative_1992}. In dimensionless form [see e.g.~\citep{leo_temporal_2010} for normalisation], and assuming anomalous dispersion, the equation reads:
\begin{equation}
\frac{\partial E}{\partial t} = \left[-1 + i(|E|^2 - \Delta) + i\frac{\partial^2}{\partial\tau^2}\right]E + S(\tau).
\label{LLE}
\end{equation}
Here $E\equiv E(t,\tau)$ is the slowly-varying envelope of the electric field in the resonator, $t$ is a slow time variable that describes the evolution of the field envelope over consecutive round trips, and $\tau$ is a fast time variable that describes the envelope's profile over a single round trip. The terms on the right-hand-side of Eq.~\eqref{LLE} respectively describe cavity losses, Kerr nonlinearity, detuning between the driving field and the closest cavity resonance ($\Delta$ is the detuning normalised to the resonance half-width), anomalous group-velocity dispersion and coherent driving, with $S(\tau)$ the envelope of the driving field. For CW driving, $S(\tau)\equiv S_0$ where $S_0$ is a constant scalar; the presence of driving field inhomogeneities is in contrast described by a driving field with explicit fast time ($\tau$) dependence.

The model described by Eq.~\eqref{LLE} can be understood to consist of two real coupled partial differential equations (PDEs) that describe the evolution of the the real ($E_r = \text{Re}[E]$) and imaginary ($E_i = \text{Im}[E]$) components of the field envelope.  To understand how CSs react to driving field inhomogeneities, we first find the steady-state ($\partial E(t,\tau)/\partial t = 0$) CS solutions of the model with homogeneous driving, $S(\tau) = S_0$. To this end, we use a multivariate Newton's method [see e.g.~\citep{kelley_solving_2003}], where the fast time is discretized into $N$ elements and the partial derivatives $\partial^2/\partial\tau^2$ computed at each grid point to transform Eq.~\eqref{LLE} into $2N$ real, coupled, ordinary differential equations (the factor of 2 comes from separation into real and imaginary parts). The system of equations can be written succinctly as
\begin{equation}
\frac{\partial \textbf{E}}{\partial t} = \textbf{f}(\textbf{E}),
\label{LLEv}
\end{equation}
where the discretized vector $\textbf{E}\in \mathbb{R}^{2N}$ contains both the real and imaginary components of the field envelope computed at all the grid points and $\textbf{f}:\mathbb{R}^{2N}\rightarrow\mathbb{R}^{2N}$ is a vector-valued function that represents the right-hand-side of Eq.~\eqref{LLE}. Newton's method iteratively finds the approximate solution to the system of equations $\textbf{f}(\textbf{E}) = \textbf{0}$ and hence the approximate steady-state solution of Eq.~\eqref{LLE}. Starting from a suitable initial guess $\textbf{E}^{(0)}$, the vector-valued function $\textbf{f}$ is approximated in the vicinity of the initial guess as the linearization
\begin{equation}
\textbf{f}(\textbf{E})\approx\textbf{f}\left(\textbf{E}^{(0)}\right) + \textbf{J}_\textbf{f}\left(\textbf{E}^{(0)}\right)\left[\textbf{E}-\textbf{E}^{(0)}\right],
\label{LLEv2}
\end{equation}
where $\textbf{J}_\textbf{f}\left(\textbf{E}^{(0)}\right)$ is the Jacobian matrix of $\textbf{f}(\textbf{E})$ evaluated at $\textbf{E}^{(0)}$. An improved approximation for the solution is then obtained by solving for the root of the right-hand-side of Eq.~\eqref{LLEv2}, yielding $\textbf{E}^{(1)} = \textbf{E}^{(0)} - \left[\textbf{J}_\textbf{f}\left(\textbf{E}^{(0)}\right)\right]^{-1}\textbf{f}\left(\textbf{E}^{(0)}\right)$. This procedure is then iteratively repeated until the solution converges to $\textbf{E}_\mathrm{s}$ for which $\textbf{f}(\textbf{E}_\mathrm{s})\approx 0$.

Once the approximate steady-state solution $\textbf{E}_\mathrm{s}$ is found, the eigenvalues and eigenvectors of the Jacobian matrix $\textbf{J}_\textbf{f}(\textbf{E}_\mathrm{s})$ provide information on the stability of the solution. In particular, the solution is dynamically stable only if none of the eigenvalues have a positive real part. Because of the translational symmetry of the system, there is always one eigenvector with zero eigenvalue; this corresponds to a \emph{neutral} (or Goldstone) mode and its existence reflects the fact that, for the case of homogeneous driving [$S(\tau) = S_0$], CSs can occupy any position along the fast time axis.

The neutral mode underpins the fact that CSs can experience time-domain motion when subject to inhomogeneous driving fields (or other forms of inhomogeneities). Considering a steady-state solution $\textbf{E}_\mathrm{s}$ with a CS centred at some temporal position $\tau_0$, we may write an inhomogeneous driving field as $S(\tau) = S_0 + P(\tau)$, where \mbox{$S_0 \equiv S(\tau_0)$} is the local driving field amplitude at the CS position and $P(\tau)$ is a (possibly complex) continuous function that represents the inhomogeneity. (Note that $P(\tau_0) = 0$ by definition.) The CS then obeys the perturbed LLE, which can be discretized into the system of equations
\begin{equation}
\frac{\partial \textbf{E}}{\partial t} = \textbf{f}(\textbf{E}) + \textbf{P},
\label{driftv}
\end{equation}
where the vector $\textbf{P}\in\mathbb{R}^{2N}$ contains both the real and imaginary components of $P(\tau)$ at all the grid points along fast time $\tau$. Assuming that the perturbation amplitude and its variations are small in the vicinity of the CS, it can be shown~\citep{maggipinto_cavity_2000} that the soliton will experience time-domain drift at a rate given by
\begin{equation}
v = \frac{d\tau_0}{dt} = \frac{\langle \textbf{v}_0|\textbf{P}\rangle}{\left\langle \textbf{v}_0\Big|\displaystyle\frac{d\textbf{E}_\mathrm{s}}{d\tau}\right\rangle},
\label{driftv2}
\end{equation}
where $\textbf{v}_0\in\mathbb{R}^{2N}$ is the left-eigenvector with zero eigenvalue of the Jacobian matrix evaluated at $\textbf{E}_\mathrm{s}$, that is, $\left[\textbf{J}_\textbf{f}(\textbf{E}_\mathrm{s})\right]^\mathrm{T}\textbf{v}_0 = 0$, and the inner product $\langle\textbf{v}_0|\textbf{P}\rangle\in\mathbb{R}$ represents a vector dot product between $\textbf{v}_0$ and $\textbf{P}$. The numerator of Eq.~\eqref{driftv2} shows that the CS drift velocity is governed by the projection of the perturbation $\textbf{P}$ on the neutral mode (here described by $\textbf{v}_0$). Recasting the discretized vectors into continuous functions so as to facilitate the analyses that will follow, we write the drift velocity as
\begin{equation}
v = \frac{d\tau_0}{dt} = \frac{\langle v_{0r}(\tau)|P_r(\tau)\rangle + \langle v_{0i}(\tau)|P_i(\tau)\rangle}{\left\langle v_{0r}\Big|\displaystyle\frac{dE_{sr}}{d\tau}\right\rangle + \left\langle v_{0i}\Big|\displaystyle\frac{dE_{si}}{d\tau}\right\rangle},
\label{driftv3}
\end{equation}
where the subscripts $r$ and $i$ refer to real and imaginary parts and the inner product now represents an integration over fast time $\tau$.

Equations~\eqref{driftv2} and~\eqref{driftv3} allow the CS drift velocity to be computed for arbitrary inhomogeneities $P(\tau)$. Here, one first finds the steady-state CS solution for a homogeneous driving amplitude $S_0$ using Newton's method, then computes the left eigenvectors $\textbf{v}_0$ of the Jacobian with zero eigenvalue, and finally evaluates the inner product to find the velocity. In what follows, we review the drift velocities that are predicted for the case of pure phase and amplitude inhomogeneity, respectively.

\subsection{Phase modulation}
For a CW driving field with pure phase modulation, we have $S(\tau) = S_0\exp[i\phi(\tau)]$, where $\phi(\tau)$ is a fast-time-dependent phase profile. Before using the normal mode theory outlined above, we note that a straightforward change of variables $E = E'\exp[i\phi(\tau)]$ in Eq.~\eqref{LLE} predicts a CS drift velocity $v_\mathrm{PM} = 2\phi'(\tau_0)$, where $\tau_0$ is the position of the CS and $\phi'(\tau) = d\phi/d\tau$~\citep{firth_optical_1996}. This same result can also be derived from simple physical arguments based on an instantaneous frequency shift caused by the time-varying phase profile $\phi(\tau)$~\citep{jang_temporal_2015}; this latter approach further highlights the linear origins of CS drift due to phase perturbations. For the sake of completeness (and as a test of the validity of the normal mode theory), we now proceed to compute the drift velocity using Eqs.~\eqref{driftv2} and~\eqref{driftv3}.

Without loss of generality, we can assume $\phi(\tau_0) = 0$ at the CS position $\tau_0$. Accordingly, in the vicinity of the soliton, we may approximate $S(\tau)\approx S_0[1+i\phi(\tau)] = S_0 + iS_0\phi(\tau)$. The perturbation $P(\tau)$ is therefore purely imaginary, with $P_i(\tau) = S_0\phi(\tau)$. Expanding $\phi(\tau)$ as a Taylor series around the soliton, $\phi(\tau)\approx \phi'(\tau_0)(\tau-\tau_0)$ where $\phi'(\tau) = d\phi/d\tau$, substitution of this expression into Eq.~\eqref{driftv3} yields:
\begin{equation}
v_\mathrm{PM} = \frac{d\tau_0}{dt} = a_\mathrm{PM}(\Delta,S_0)\phi'(\tau_0),
\end{equation}
where the coefficient $a_\mathrm{PM}(\Delta,S_0)$ is given by
\begin{equation}
a_\mathrm{PM}(\Delta,S_0) = S_0\frac{\langle v_{0i}(\tau-\tau_0)|\tau-\tau_0\rangle}{\left\langle v_{0r}\Big|\displaystyle\frac{dE_{sr}}{d\tau}\right\rangle + \left\langle v_{0i}\Big|\displaystyle\frac{dE_{si}}{d\tau}\right\rangle},
\label{apm}
\end{equation}
where we used $\tau-\tau_0$ as the argument for $v_{0i}$ to highlight that the neutral mode is centred at the CS position $\tau_0$. Note also the multiplicative prefactor $S_0$.

\begin{figure}[!t]
 \centering
  \includegraphics[width = 0.9\textwidth, clip=true]{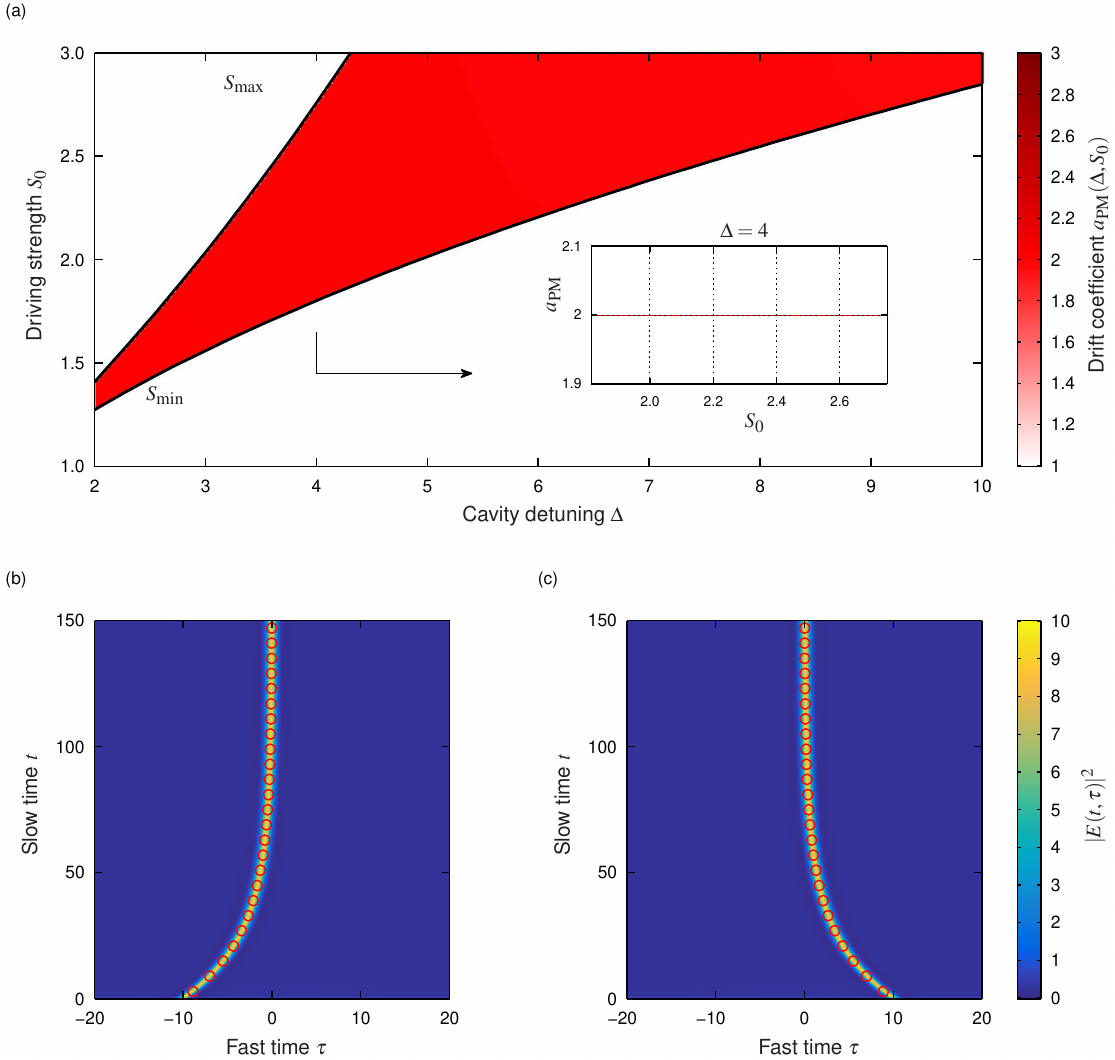}
 \caption{(a) Pseudocolor plot showing the phase drift coefficient $a_\mathrm{PM}(\Delta,S_0)$ as a function of the cavity detuning $\Delta$ and driving amplitude $S_0$. As can be seen, the coefficient $a_\mathrm{PM}(\Delta,S_0) = 2$ for all $\Delta$ and $S_0$ [see also inset for the curve $a_\mathrm{PM}(\Delta = 4,S_0)$ in more detail]. The black curves approximate the maximum ($S_\mathrm{max}$) and minimum ($S_\mathrm{min}$) driving amplitudes required for CS existence. (b) and (c) show illustrative CS dynamics  in the presence of driving field phase inhomogeneities obtained via direct integration of Eq.~\eqref{LLE}. Both illustrations use $X = \Delta = 4$ and a parabolic phase profile $\phi(\tau)=-\phi_0\tau^2/\sigma^2$ with $\sigma = 10$ and $\phi_0=1~\mathrm{rad}$. Red circles show theoretically predicted trajectories as described in the text. The bottom colorbar is common to (b) and (c).}
 \label{fig2}
\end{figure}

Figure~\ref{fig2}(a) shows the coefficient $a_\mathrm{PM}(\Delta,S_0)$ as computed for a range of detunings $\Delta$ and driving amplitudes $S_0$ using Eq.~\eqref{apm}. As can be seen, the coefficient $a_\mathrm{PM}$ assumes the constant value $a_\mathrm{PM}(\Delta,S_0) = 2$ for all detunings $\Delta$ and driving amplitudes $S_0$. Accordingly, the normal mode theory indeed produces the expected drift velocity $v_\mathrm{PM} = 2\phi'(\tau_0)$.

A direct consequence of the result $v_\mathrm{PM} = 2\phi'(\tau_0)$ is that a CS will always drift along (against) an increasing (decreasing) phase gradient. In other words, the solitons will always move towards the maxima of the phase modulation, where $\phi'(\tau) = 0$ and hence $v_\mathrm{PM} = 0$. These maxima correspond to robust trapping sites: if a perturbation shifts the CS away from the maximum, the soliton will simply drift back. In contrast, whilst the CS velocity is also zero at the minima of the phase modulation, these correspond to unstable equilibria that cannot act as trapping sites in practice.

Figure~\ref{fig2}(b) and (c) show illustrative dynamics of phase-modulation induced CS drift, obtained via numerical integration of Eq.~\eqref{LLE}; similar simulation results have been reported in a range of studies~\citep{jang_temporal_2015, luo_spontaneous_2015, taheri_soliton_2015, lobanov_harmonization_2016}. Here, Figs.~\ref{fig2}(b) and (c) respectively show the spatiotemporal evolution of a CS initially excited at $\tau_0 = -10$ and $\tau_0 = +10$ under the influence of a driving field with parabolic phase profile $\phi(\tau)=-\phi_0\tau^2/\sigma^2$. (For other simulation parameters, see figure caption.) As can be seen, the soliton is attracted towards -- and traps to -- the maximum of the phase perturbation at $\tau = 0$. Also shown as the red circles in Figs.~\ref{fig2}(b) and (c) are the theoretically predicted CS trajectories $\tau_0(t)$ obtained by directly integrating the differential equation $d\tau_0/dt = 2\phi'(\tau)$; for the particular phase profile chosen, this yields $\tau_0(t) = \tau_0(0)\exp(-4\phi_0 t/\sigma^2)$. We see very good agreement between the theoretical prediction and numerically simulated dynamics. Experimental realisations of phase modulation trapping of temporal CSs will be surveyed in Section~\ref{PMexperiments}.

\subsection{Amplitude modulation}
We now consider the case of a driving field that exhibits pure amplitude inhomogeneities. The underlying physics is more complicated than for the case of pure phase inhomogeneities~\citep{hendry_spontaneous_2018}, and there is no known straightforward method to compute the CS drift velocity for amplitude inhomogeneities: the normal mode theory is necessary. To proceed, we approximate the driving field in the vicinity of the soliton as a first-order Taylor series expansion: \mbox{$S(\tau)\approx S_0 + (\tau-\tau_0)S'(\tau_0)$}, where $S'(\tau_0) = dS/d\tau|_{\tau_0}$. Assuming pure amplitude inhomogeneity, the perturbation $P(\tau)$ is real with $P_\mathrm{r}(\tau) = (\tau-\tau_0)S'(\tau_0)$. Substituting this expression into Eq.~\eqref{driftv3} yields:
\begin{equation}
v_\mathrm{AM} = \frac{d\tau_0}{dt} = a_\mathrm{AM}(\Delta,S_0)S'(\tau_0),
\end{equation}
where the coefficient $a_\mathrm{AM}(\Delta,S_0)$ is given by
\begin{equation}
a_\mathrm{AM}(\Delta,S_0) = \frac{\langle v_{0r}(\tau-\tau_0)|\tau-\tau_0\rangle}{\left\langle v_{0r}\Big|\displaystyle\frac{dE_{sr}}{d\tau}\right\rangle + \left\langle v_{0i}\Big|\displaystyle\frac{dE_{si}}{d\tau}\right\rangle}.
\label{aam}
\end{equation}
Note that, in contrast to the corresponding expression for phase modulation [see Eq.~\eqref{apm}], the coefficient $a_\mathrm{AM}$ is obtained by projecting the \emph{real} part of the neutral mode along a linear fast time variation, and the result is \emph{not} multiplied by the local driving field amplitude $S_0 = S(\tau_0)$.

\begin{figure}[!t]
 \centering
  \includegraphics[width = 0.9\textwidth, clip=true]{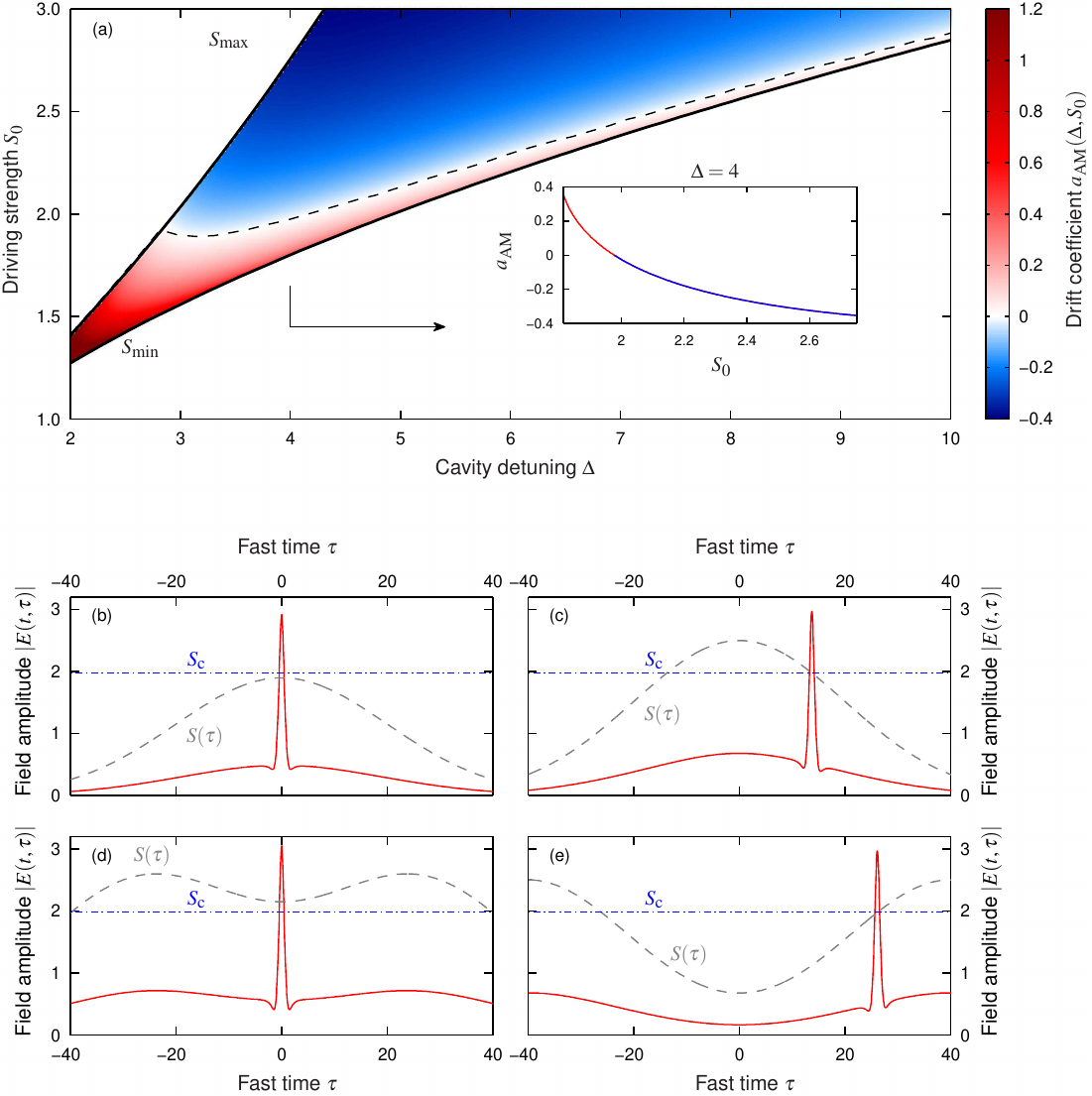}
 \caption{(a) Pseudocolor plot showing the amplitude drift coefficient $a_\mathrm{AM}(\Delta,S_0)$ as a function of the cavity detuning $\Delta$ and driving amplitude $S_0$. Inset shows the curve $a_\mathrm{AM}(\Delta = 4,S_0)$, the black solid curves approximate the maximum ($S_\mathrm{max}$) and minimum ($S_\mathrm{min}$) driving amplitudes required for CS existence, and the black dashed curve highlights the points in the parameter space where $a_\mathrm{AM}(\Delta,S_0) = 0$. (b)--(e) Steady-state CS profiles for different driving field amplitude profiles. (b) and (c) consider a Gaussian driving field $S(\tau) = S_\mathrm{g}e^{-\tau^2/\sigma^2}$ with duration $\sigma = \sqrt{800}$ and peak amplitude (b) $S_\mathrm{g} = 1.9$  and (c) $S_\mathrm{g} = 2.5$. (d) and (e) consider a driving field that consists of a superposition of two Gaussian profiles ($\sigma=\sqrt{800}$, $S_\mathrm{g}=2.5$) separated from one another by (d) $\Delta \tau = 26$ and (e) $\Delta \tau = 40$. The blue dash-dotted lines indicate the critical driving amplitude $S_\mathrm{c}\approx 1.98$. Panels (a)--(c) adapted from~\citep{hendry_spontaneous_2018}.}
 \label{fig3}
\end{figure}

Figure~\ref{fig3}(a) shows the coefficient $a_\mathrm{AM}(\Delta,S_0)$ as computed for a range of detunings $\Delta$ and driving amplitudes $S_0$ using Eq.~\eqref{apm}. In stark contrast with the result obtained for pure phase modulation [see Fig.~\ref{fig2}], $a_\mathrm{AM}(\Delta,S_0)$ depends nontrivially on both the detuning and the driving amplitude. We see that, for large detunings $\Delta>3$, there exists a single ``critical'' driving amplitude $S_\mathrm{c}$ such that $a_\mathrm{AM}(\Delta,S_\mathrm{c}) = 0$.  Points $\tau_\mathrm{c}$ along the fast time that satisfy $S(\tau_\mathrm{c}) = S_\mathrm{c}$  correspond to stable equilibria towards which CSs are attracted -- and subsequently trapped -- to. Importantly, these points do not in general correspond to the extrema of the driving field: the CS may be trapped at the edge of an amplitude inhomogeneity if that is where the critical value $S_\mathrm{c}$ is reached~\citep{hendry_spontaneous_2018}. It is worth noting that the extrema of the driving field are also equilibria [$S'(\tau) = 0$ implies that $v_\mathrm{AM} = 0$], but these are only stable if there is no monotonic route to the critical value.

Figures~\ref{fig3}(b)--(e) show steady-state CS profiles obtained by direct numerical simulations of Eq.~\eqref{LLE} that illustrate the different trapping dynamics in the presence of amplitude inhomogeneities. The simulations consider a constant detuning of $\Delta = 4$ (for which the critical value $S_\mathrm{c}= 1.98$) but different driving profiles. Figures~\ref{fig3}(b) and (c) use a Gaussian driving field $S(\tau) = S_\mathrm{g}e^{-\tau^2/\sigma^2}$ with two different peak amplitudes [see Figure caption for details]. When the peak amplitude $S_\mathrm{g}<S_\mathrm{c}$ [Fig.~\ref{fig3}(b)], we see that the CS is attracted to the peak of the driving field -- that being the point closest to the critical value $S_\mathrm{c}$. In contrast, when $S_\mathrm{g}>S_\mathrm{c}$, we see that the CS is attracted to the edge of the driving field to a point $\tau_\mathrm{c}$ where $S(\tau_\mathrm{c}) = S_\mathrm{c}$ [Fig.~\ref{fig3}(c)]. Figures~\ref{fig3}(d) and (e) consider a driving profile comprised of a superposition of two Gaussian pulses offset from one another [see Figure caption for details] so as to realise a local dip in the driving field. When the driving strength at the dip is greater than $S_\mathrm{c}$, the CS is attracted to that point [Fig.~\ref{fig3}(d)]. In contrast, when the driving strength at the dip is smaller than $S_\mathrm{c}$, the CS is attracted to the point where the driving field attains the critical value (or a local maximum if the critical value is never attained).


\subsection{Desynchronization and locking range}
The analysis presented above has assumed that the phase or amplitude inhomogeneity is perfectly synchronised with the natural cavity round trip time (reciprocal of the cavity free-spectral range). This is reflected by the fact that the driving term in Eq.~\eqref{LLE} is not dependent on the slow time $t$. In contrast, in the presence of desynchronisation, the externally applied inhomogeneity is advanced or delayed with respect to the intracavity field each round trip, such that $S(\tau)\rightarrow S(\tau+d\times t)$, where $d$ represents the amount of drift. By changing into a reference frame where the pump is stationary, $\tau\rightarrow\tau - d\times t$, the LLE Eq.~\eqref{LLE} develops a convective drift term~\citep{coen_convection_1999,parra-rivas_effects_2014, hendry_impact_2019}:
\begin{equation}
\frac{\partial E}{\partial t} = \left[-1 + i(|E|^2 - \Delta) -d\frac{\partial}{\partial\tau}+ i\frac{\partial^2}{\partial\tau^2}\right]E + S(\tau).
\label{LLEd}
\end{equation}

As discussed in~\citep{parra-rivas_effects_2014, hendry_impact_2019}, the pump-cavity desynchronization gives rise to a relative drift between the CS and the driving field which is in addition to the motion induced by phase or amplitude inhomogeneities. Therefore, using $a_\mathrm{PM}(S_0,\Delta) = 2$, the total CS drift velocity becomes
\begin{equation}
v =
\left\{
	\begin{array}{ll}
		2\phi'(\tau_0) + d  & \mbox{for phase modulation}, \vspace{5pt} \\
		a_\mathrm{AM}(\Delta,S_0)S'(\tau_0) + d & \mbox{for amplitude modulation}.
	\end{array}
\right.
\label{vdrift}
\end{equation}
It should be clear that desynchronization shifts the CS locking positions to new values where $v = 0$; this shifting can also affect the number of CS configurations that can be sustained~\citep{parra-rivas_effects_2014, hendry_impact_2019}. Of course, if the desynchronization is too large, the equation $v = 0$ may not possess any real roots; in this case, the soliton will be unlocked, i.e., it will not be trapped to the inhomogeneity at all, but will rather circulate at its own natural repetition rate. The range of desynchronizations that can be compensated for by the inhomogeneity is known as the locking range and is given by $d_\mathrm{max} = 2|\phi'(\tau_0)|_\mathrm{max}$ for phase inhomogeneities and $d_\mathrm{max} = |a_\mathrm{AM}(\Delta,S_0)S'(\tau_0)|_\mathrm{max}$ for amplitude inhomogeneities. For the sake of completeness, we present below the corresponding dimensional expressions for the locking range in terms of the frequency mismatch $\Delta f = \text{FSR}-f_\mathrm{H}$, where $\text{FSR}$ is the cavity free-spectral range (reciprocal of the natural round trip time) and $f_\mathrm{H}$ is the repetition frequency of the driving field inhomogeneity ($f_\mathrm{H}\sim\text{FSR}$):
\begin{equation}
\Delta f_\mathrm{max} \approx
\left\{
	\begin{array}{ll}
		|\beta_2|L \,\text{FSR}^2|\phi'(\tau)|_\mathrm{max} & \mbox{for phase modulation,} \vspace{10pt} \\
		\frac{\displaystyle|\beta_2|L\, \text{FSR}^2}{2}\sqrt{\displaystyle\frac{\gamma L\theta\mathcal{F}^3}{\pi^3}}\left|b_\mathrm{AM}(\delta_0,E_\mathrm{in})E_\mathrm{in}'(\tau)\right|_\mathrm{max} & \mbox{for amplitude modulation.}
	\end{array}
\right.
\label{lockingr}
\end{equation}
Here $\beta_2$ is the group-velocity dispersion coefficient (with units of $\mathrm{s^2m^{-1}}$), $L$ is the roundtrip length of the resonator, $\gamma$ is the Kerr nonlinearity coefficient (with units of $\mathrm{W^{-1}m^{-1}}$), $\theta$ is the power transmission coefficient of the coupler used to inject the driving field into the resonator, $E_\mathrm{in}(\tau)$ is the complex electric field envelope of the driving field (with units of $\mathrm{W^{1/2}}$), $\mathcal{F}$ is the finesse of the resonator, $\delta_0 = \pi\Delta/\mathcal{F}$ is the phase detuning between the driving field and the closest cavity resonance, and the coefficient $b_\mathrm{AM}(\delta_0,E_\mathrm{in}) = a_\mathrm{AM}(\delta_0\mathcal{F}/\pi,E_\mathrm{in}[\gamma L\theta\mathcal{F}^3/\pi^3]^{1/2})$.

Equations~\eqref{lockingr} show that that the ``strength'' of the phase modulation trapping, and hence the attainable locking range, does not depend on the nonlinearity coefficient. This reflects the linear origins of CS motion due to phase inhomogeneities. In contrast, CS motion in the presence of amplitude inhomogeneities is intrinsically a nonlinear process~\citep{hendry_spontaneous_2018}, and the locking range in that case is indeed found to depend on the nonlinearity coefficient $\gamma$. Because of the complicated functional form of the amplitude drift coefficient $a_\mathrm{AM}(\Delta,S_0)$, it is not straightforward to quantitatively compare its strength with respect to trapping due to phase inhomogeneities. Dedicated research is needed to compare the two processes.

\section{Experimental results}
The section above summarised the salient theoretical details pertinent to drift and trapping of temporal Kerr CSs due to inhomogeneous driving fields. Here in this Section we review some of the key experimental results that confirm, apply, and extend upon the theoretical developments.

\subsection{Phase modulation experiments}
\label{PMexperiments}
\begin{figure}[!b]
 \centering
  \includegraphics[width = \textwidth, clip=true]{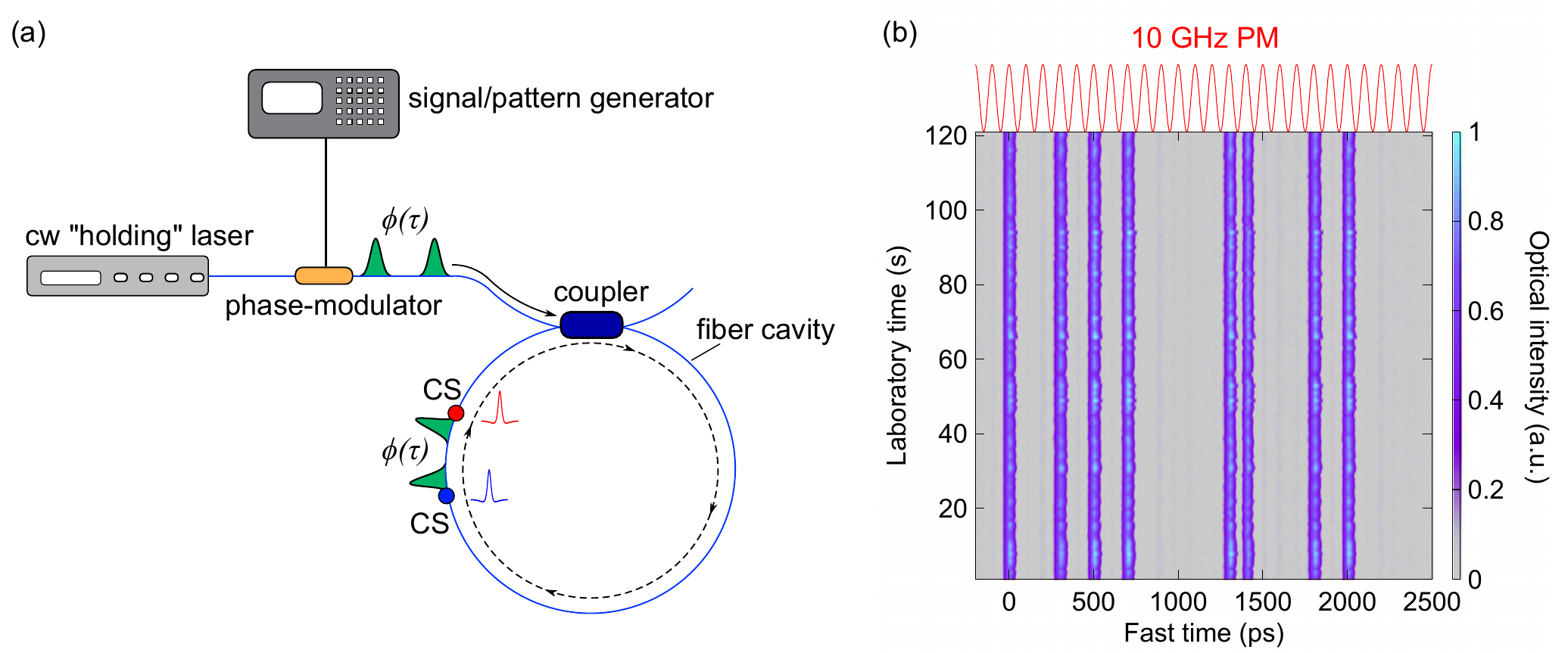}
 \caption{(a) Schematic illustration of a CS-supporting fibre ring resonator driven with a phase modulated field. (b) Experimental results, showing the trapping of eight temporal CSs to the peaks of a 10~GHz phase modulation profile (inset). Figure adapted from~\citep{jang_temporal_2015}. }
 \label{fig4}
\end{figure}
The fact that phase modulations imposed on the cavity driving beam can be used to trap CS at desired positions has been demonstrated first in the context of spatial CSs in diffractive microcavities~\citep{pedaci_positioning_2006}. In the context of temporal CSs (which are at the focus of our Article), the first experimental demonstrations were reported in 2015 by Jang et al.~\citep{jang_temporal_2015}. The authors excited CSs in a macroscopic (100-m-long) optical fibre ring resonator and used a 10~GHz electro-optic phase modulator to imprint a desired phase profile on the otherwise CW driving field [see Fig.~\ref{fig4}(a)]. Figure~\ref{fig4}(b) shows an illustrative experimental result of eight CS circulating the resonator simultaneously, each trapped to the maximum of an underlying 10~GHz sinusoidal phase modulation. It must be emphasised that, in the absence of phase modulation, the solitons would undergo spontaneous motion due to long-range interactions mediated by forward Brillouin scattering~\citep{jang_ultraweak_2013}; the fact that no such motion occurs during the two-minute experimental measurement shown in Fig.~\ref{fig4}(b) attests to the solitons being robustly trapped to the sinusoidal phase modulation. In addition to demonstrating the general trapping potential of phase modulation, the experiments performed by Jang et al. also meticulously confirmed that, in the presence of phase modulations, the expression $v_\mathrm{PM} = 2\phi'(\tau_0)$ derived above indeed provides a correct prediction for the CS drift velocity.

Following the demonstration of Jang et al., phase modulation has been widely used to trap CSs in several fibre ring resonator experiments~\citep{wang_addressing_2018,xu_spontaneous_2020}, allowing e.g. for the realisation of a CS-based all-optical buffer operating at 10~GHz~\citep{jang_all-optical_2016}. Phase modulation has also been used in microresonator experiments to facilitate the formation of single-soliton frequency combs states that are locked to an external RF signal~\citep{cole_kerr-microresonator_2018}. In particular, in 2018, Cole et al. used a 9.5~mm silica wedge resonator with 22~GHz free-spectral range, driven with a CW laser phase-modulated at the fundamental FSR, to demonstrate an operating regime where only the single-soliton state is permitted -- at a position that coincides with the phase modulation maximum. The authors also showed that the phase modulation trapping allows for the systematic control and stabilization of the soliton repetition rate ($f_\mathrm{rep}$).  In a very recent study, the phase modulation techniques demonstrated by Cole et al. were used to realise a frequency-stabilised coherent soliton microcomb that was subsequently demonstrated to allow for direct atomic frequency comb spectroscopy of rubidium~\citep{stern_direct_2020}.


\subsection{Amplitude modulation experiments}
Amplitude modulated driving fields are inherent to experiments in spatially diffractive resonators due to the impossibility of realising a perfect plane wave (rather, the driving fields tend to have broad Gaussian profiles). In contrast, in the context of dispersive resonators, CW lasers naturally provide temporally homogeneous driving fields with only minor amplitude inhomogeneities. However, as noted in the introduction of this Article, CW driving suffers from certain shortcomings that can be mitigated by employing driving fields that are not homogeneous. Numerous studies have in particular considered resonators that are driven with a train of pulses whose repetition rate is carefully synchronised to the free-spectral range of the resonator, thus ensuring that each round trip a new driving pulse is coherently added on the circulating intracavity field.

Pulsed driving has been widely used in experiments involving CSs in macroscopic fibre ring resonators~\citep{anderson_observations_2016, anderson_coexistence_2017, wang_stimulated_2018, nielsen_coexistence_2019, xu_spontaneous_2020, dong_stretched-pulse_2020}. These studies typically use flat-top nanosecond pulses that are much longer than the picosecond-scale CSs. Such nanosecond driving fields can be considered ``quasi-CW'' in that the amplitude inhomogeneity does not play any significant role for CSs residing in the flat-top portion of the field. The benefit compared to pure CW driving is, however, the possibility of reaching much greater driving amplitudes: assuming identical amplification schemes, the \emph{peak} power attainable for a pulse train is larger than the power attainable for a CW field by a factor given by the reciprocal of the pulse train's duty cycle. Another benefit of quasi-CW driving in the context of fibre ring resonator experiments is the fact that such driving is immune to stimulated Brillouin scattering, which is known to hinder experiments in the pure CW regime.

\begin{figure}[!t]
 \centering
  \includegraphics[width = \textwidth, clip=true]{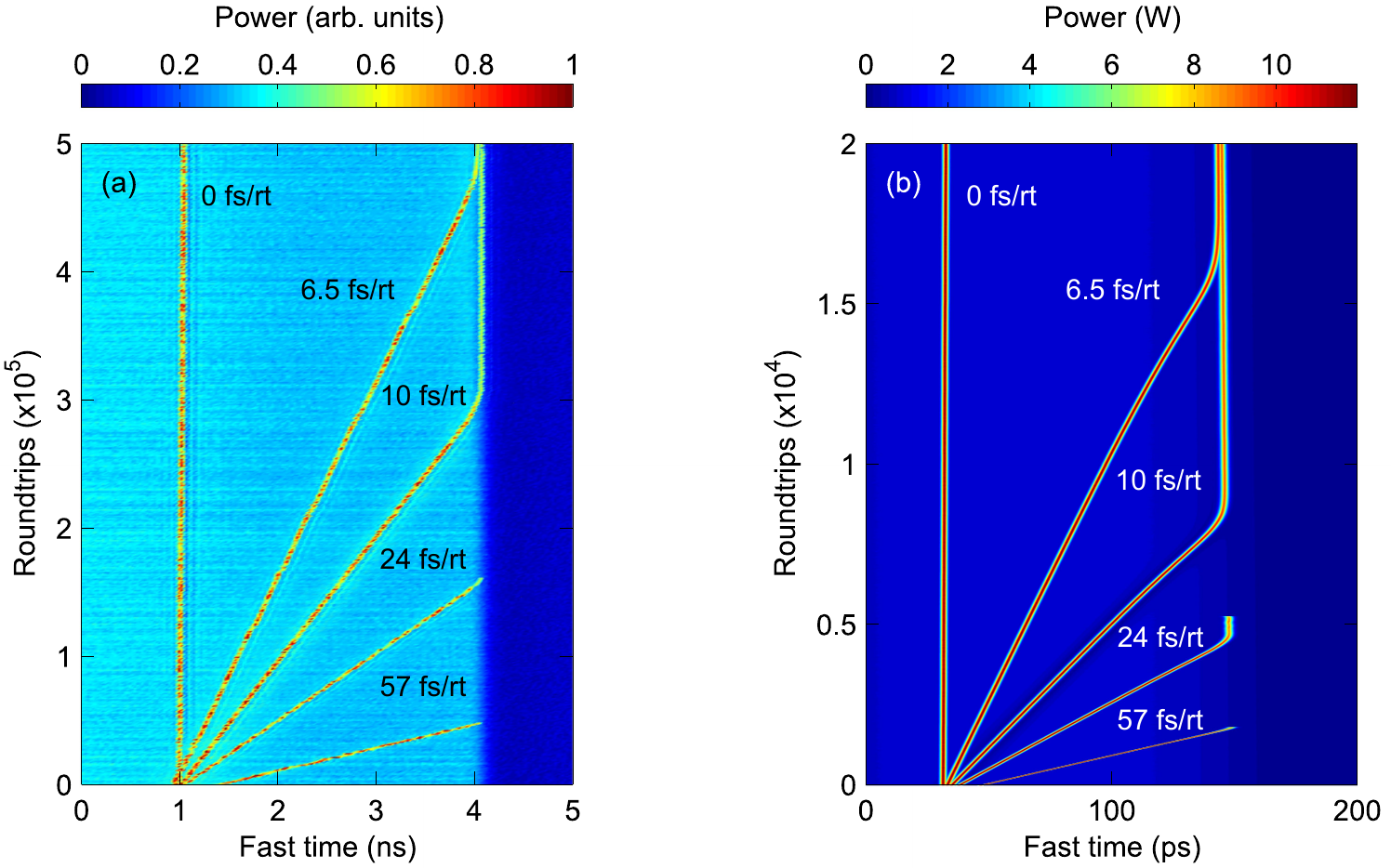}
 \caption{Assortment of (a) experimental measurements and (b) numerical simulation results for different desynchronizations that show how CSs can be trapped at the edge of amplitude inhomogeneities. The driving field is a nanosecond-scale flat-top pulse. Note that the trajectories in (a) and (b) correspond to independent realisations that have been superimposed on the same background to facilitate visualisation. Adapted from~\citep{wang_temporal_2018}.}
 \label{fig5}
\end{figure}

The use of quasi-CW pulsed driving was initially motivated by the access to larger driving powers, but it was also quickly realised that the amplitude inhomogeneity allows for CS trapping. In particular, residual desynchronization between the driving pulse train and the cavity free-spectral range causes the CSs to drift towards the leading or trailing edge of the driving field, where the solitons were found to be robustly trapped over a wide range of desynchronizations~\citep{wang_temporal_2018, nielsen_coexistence_2019}. Figure~\ref{fig5} shows an example of such dynamics, with Figs.~\ref{fig5}(a) and (b) showing experimental and simulated results, respectively. These results superimpose five different realisations where the desynchronization was adjusted from 0~fs per round trip to 57~fs per round trip, with a peak driving amplitude above the critical value. We clearly see how the CS can become trapped at the driving pulse edge for sufficiently small desynchronizations.

More recent experiments have also demonstrated pulsed driving to represent a viable technique to realise temporal CSs (and corresponding optical frequency combs) in high-Q microresonators. In these experiments (and in contrast to the fibre ring experiments described above), the pulses used are very short (of the order of picosecond), thus shifting truly beyond the paradigm of (quasi-)CW driving. The first demonstration of this nature was achieved by Obrzud et al., who realised CS optical frequency combs in a 1-cm-long fibre-based Fabry-Perot microresonator driven with pulses as short as 2.1~ps~\citep{obrzud_temporal_2017}. The driving pulse train was generated using an electro-optic comb generation scheme~\citep{kobayashi_highrepetitionrate_1972, kobayashi_optical_1988, beha_electronic_2017}: CW laser light was passed through a cascaded sequence of phase and amplitude modulators and compressed into picosecond pulses by a chirped fibre-Bragg grating. It was found that temporal CSs with 200-fs-duration could form on top of the resonantly-enhanced, 2~ps intracavity pump pulses. The authors also demonstrated that the soliton repetition rate locks to the repetition of the driving pulse train, and also examined the range of desynchronizations over which solitons can be sustained. Subsequent numerical modelling has shown that stimulated Raman scattering plays a key role in defining that range~\citep{hendry_impact_2019}.

Following the pioneering work by Obrzud et al., similar electro-optically generated pulse trains have been used to realise CSs in a range of different microresonator architectures, including silicon nitride ring resonators~\citep{anderson_photonic_2020} and integrated silica resonators~\citep{xu_harmonic_2020}. These studies have also shown that CSs can be generated when the resonator is driven at rational harmonic fractions of the cavity FSR; one such scheme has also been successfully used to realise a microresonator CS-based ``astrocomb'' for the detection of extra-solar planets~\citep{obrzud_microphotonic_2019}. It is worth noting that, even though the input electro-optically generated pulse train already corresponds to a frequency comb, the CS that is generated typically has considerably shorter duration (and hence broader bandwidth) than the input. Moreover, it has been beautifully demonstrated that the soliton trapping dynamics suppress noise transfer from the input pulse train to the CS -- the soliton effectively exhibits lesser timing jitter than the input pump pulse~\citep{brasch_nonlinear_2019, anderson_photonic_2020}. This result can be readily appreciated from Eq.~\eqref{vdrift} by first writing $a_\mathrm{AM}(S_0,\Delta)S'(\tau_0) = f(\tau_0)$ to obtain
\begin{equation}
v=\frac{d\tau_0}{dt} = f(\tau_0) + d(t),
\end{equation}
where the time-dependent desynchronization $d(t) = d_0 + \varepsilon(t)$ includes a constant component $d_0$ and a random jitter term $\varepsilon(t)$ that is assumed small ($\varepsilon\ll d_0$). Assuming further that $f(\tau_0)$ varies slowly, we may linearize the term around the average trapping position -- taken to be at $\tau_0 = 0$ without loss of generality -- to yield
\begin{equation}
v = \frac{d\tau_0}{dt} = f'(0)\tau_0+\varepsilon(t),
\end{equation}
where $f'(0) = df/d\tau_0|_0$ and we used the fact that $f(0) + d_0 = 0$ by virtue of \mbox{$\tau_0 = 0$} being an equilibrium. Since $\tau_0 = 0$ corresponds to a stable equilibrium, we have $f'(0) < 0$. Considering harmonic jitter $\varepsilon(t) = \varepsilon_0\exp[i\Omega t]$ and using the ansatz $\tau_0(t) = T(\Omega)\exp[i\Omega t]$, it is straightforward to derive the frequency response~\citep{brasch_nonlinear_2019}
\begin{equation}
Z(\Omega) = \frac{|T(\Omega)|^2}{|\varepsilon_0|^2} = \frac{1}{|f'(0)|^2 + \Omega^2},
\end{equation}
revealing a Lorentzian low-pass filter with cut-off frequency of $2|f'(0)|$. In dimensional units, the cut-off frequency $\Delta f_\mathrm{cutoff, AM}$ can be written (with units of Hz) as:
\begin{equation}
\Delta f_\mathrm{cutoff, AM} = \frac{|\beta_2| L}{2\pi}\text{FSR}\sqrt{\frac{\gamma L \theta\mathcal{F}^3}{\pi^3}}|g'(0)|,
\label{cutoff}
\end{equation}
where $g'(0)$ is the (dimensional) fast time derivative of the function \mbox{$g(\tau) = b_\mathrm{AM}(\delta_0,E_\mathrm{in})E_\mathrm{in}'(\tau)$} evaluated at the mean soliton trapping position (assumed to be $\tau = 0$) and all the other parameters are as defined in Section~\ref{theorysec}. Unexpectedly, Eq.~\eqref{cutoff} shows that a small cutoff frequency can be obtained with parameters that are also associated with a small trapping strength [c.f. Eq.~\eqref{lockingr}]: if the soliton is only weakly coupled to the inhomogeneity, it will not react significantly to jitter in the inhomogeneity. We also note that the analysis above can be readily repeated for the case of phase modulation trapping, yielding the dimensional cutoff frequency:
\begin{equation}
\Delta f_\mathrm{cutoff, PM} = \frac{|\beta_2| L}{\pi}\text{FSR}|\phi''(0)|.
\label{cutoff2}
\end{equation}

Arguably one of the most important outcomes of the use of short pulsed driving to generate CSs is the fact that the technique has made it possible to realise CSs and corresponding frequency combs in a wider range of systems. Particularly noteworthy is the fact that pulsed driving has allowed temporal CSs to be generated and observed in a system that is fundamentally different from earlier waveguide-type resonator implementations, namely a free-space femtosecond enhancement cavity~\citep{lilienfein_temporal_2019}. Specifically, considering a free-space bowtie resonator containing a 1-mm-thick sapphire plate to provide a Kerr nonlinearity, driven with a fully-stabilized mode-locked laser delivering 350~fs pulses, Lilienfein et al. demonstrated the generation of CSs as short as 37~fs. The authors' measurements showed that, by leveraging CSs, it is possible to obtain peak-power enhancements an order of magnitude larger than previously demonstrated in femtosecond enhancement cavities with similar pulse durations. Compounded by the solitons' noise suppression characteristics, the results underline the potential usefulness of temporal CSs in an entirely new spectrum of applications~\citep{brasch_nonlinear_2019}, including high-order harmonic generation~\citep{gohle_frequency_2005} and photoemission spectroscopy~\citep{saule_high-flux_2019}.

\section{Conclusions and future outlook}
The use of driving fields with phase or amplitude inhomogeneities offers a number of advantages for the study and application of temporal Kerr cavity solitons, including the ability to implement robust soliton trapping and control schemes. In this Article, we have reviewed the salient theoretical descriptions of CS dynamics in the presence of inhomogeneous driving fields, and surveyed relevant experimental literature.

We envisage that inhomogeneous driving fields will continue to gain in popularity in CS-related studies, stimulating significant future research. For example, the fact that pulsed driving gives access to very large (peak) power levels is likely to further expand the systems in which CSs and related frequency combs can be generated, paving the way for new applications. The knowledge generated over the past few years also provides an avenue for optimising a given CS control parameter (such as locking range or nonlinear filtering bandwidth), which is likely to allow inhomogeneous driving fields to further enhance the performance of existing applications. In this context, it is worth noting that studies so far have focussed on phase or amplitude inhomogeneities in isolation: we expect that judiciously engineered combination of the two is likely to be key for future optimisation.

It is also important to note that our understanding of CS control and trapping with driving field inhomogeneities is so far built around situations where higher-order effects -- such as stimulated Raman scattering or higher-order dispersion -- are negligible. Considerable research is needed to understand how the underlying physics change when higher-order effects play a significant role. For example, recent experiments have shown signatures of CSs that can exist under conditions of weak normal dispersion driving~\citep{li_experimental_2020,anderson_zero-dispersion_2020}; in this regime, third-order dispersion plays a dominant role and is expected to alter the CS trapping dynamics. Similarly, the response of CSs to driving field inhomogeneities in systems with multiple interacting spatial or polarization mode families~\citep{lucas_spatial_2018, nielsen_coexistence_2019, xu_spontaneous_2020} -- or coupled cavity configurations~\citep{xue_super-efficient_2019,kim_turn-key_2019,helgason_dissipative_2020} -- represents an interesting area of further research. Finally, we envisage that the phase and amplitude trapping of other localized structures -- such as dark solitons~\citep{xue_mode-locked_2015,fulop_high-order_2018} or platicons~\citep{lobanov_frequency_2015} -- that can manifest themselves in Kerr resonators is also likely to draw research attention in the coming years.

\section*{Acknowledgements}
\noindent We acknowledge financial support from the Marsden Fund and the Rutherford Discovery Fellowships of the Royal Society of New Zealand. We would also like to thank Dr Yadong Wang for providing data and codes to plot Figure~\ref{fig5} and Mr Ian Hendry for useful discussions.

\bibliographystyle{apa}


\begin{thebibliography}{}

\bibitem[\protect\astroncite{Ackemann et~al.}{2009}]{ackemann_chapter_2009}
Ackemann, T., Firth, W., and Oppo, G.-L. (2009).
\newblock  {Fundamentals} and {Applications} of {Spatial}
  {Dissipative} {Solitons} in {Photonic} {Devices}.
\newblock \emph{Adv. Atom. Mol. Opt. Phys.} 57:323.

\bibitem[\protect\astroncite{Akhmediev and
  Ankiewicz}{2008}]{akhmediev_dissipative_2008}
Akhmediev, N. and Ankiewicz, A. (2008).
\newblock {\em Dissipative {Solitons}: {From} {Optics} to {Biology} and
  {Medicine}}, volume 751 of {\em Lecture {Notes} in {Physics}}.
\newblock Springer Berlin Heidelberg, Berlin, Heidelberg.

\bibitem[\protect\astroncite{Anderson
  et~al.}{2016}]{anderson_observations_2016}
Anderson, M., Leo, F., Coen, S., Erkintalo, M., and Murdoch, S.~G. (2016).
\newblock Observations of spatiotemporal instabilities of temporal cavity
  solitons.
\newblock {\em Optica}, 3(10):1071.

\bibitem[\protect\astroncite{Anderson et~al.}{2017}]{anderson_coexistence_2017}
Anderson, M., Wang, Y., Leo, F., Coen, S., Erkintalo, M., and Murdoch, S.~G.
  (2017).
\newblock Coexistence of {Multiple} {Nonlinear} {States} in a {Tristable}
  {Passive} {Kerr} {Resonator}.
\newblock {\em Phys. Rev. X}, 7(3):031031.

\bibitem[\protect\astroncite{Anderson et~al.}{2020a}]{anderson_photonic_2020}
Anderson, M.~H., Bouchand, R., Liu, J., Weng, W., Obrzud, E., Herr, T., and
  Kippenberg, T.~J. (2020a).
\newblock Photonic chip-based resonant supercontinuum.
\newblock {\em arXiv:1909.00022 [physics]}.

\bibitem[\protect\astroncite{Anderson
  et~al.}{2020b}]{anderson_zero-dispersion_2020}
Anderson, M.~H., Lihachev, G., Weng, W., Liu, J., and Kippenberg, T.~J.
  (2020b).
\newblock Zero-dispersion {Kerr} solitons in optical microresonators.
\newblock {\em arXiv:2007.14507 [physics]}.

\bibitem[\protect\astroncite{Bao et~al.}{2014}]{bao_nonlinear_2014}
Bao, C., Zhang, L., Matsko, A., Yan, Y., Zhao, Z., Xie, G., Agarwal, A.~M.,
  Kimerling, L.~C., Michel, J., Maleki, L., and Willner, A.~E. (2014).
\newblock Nonlinear conversion efficiency in {Kerr} frequency comb generation.
\newblock {\em Opt. Lett.}, 39(21):6126--6129.

\bibitem[\protect\astroncite{Barland et~al.}{2002}]{barland_cavity_2002}
Barland, S., Tredicce, J.~R., Brambilla, M., Lugiato, L.~A., Balle, S.,
  Giudici, M., Maggipinto, T., Spinelli, L., Tissoni, G., Kn{\"o}dl, T.,
  Miller, M., and J{\"a}ger, R. (2002).
\newblock Cavity solitons as pixels in semiconductor microcavities.
\newblock {\em Nature}, 419(6908):699--702.

\bibitem[\protect\astroncite{Beha et~al.}{2017}]{beha_electronic_2017}
Beha, K., Cole, D.~C., Del{\textquoteright}Haye, P., Coillet, A., Diddams,
  S.~A., and Papp, S.~B. (2017).
\newblock Electronic synthesis of light.
\newblock {\em Optica}, 4(4):406--411.

\bibitem[\protect\astroncite{Brasch et~al.}{2016}]{brasch_photonic_2016}
Brasch, V., Geiselmann, M., Herr, T., Lihachev, G., Pfeiffer, M. H.~P.,
  Gorodetsky, M.~L., and Kippenberg, T.~J. (2016).
\newblock Photonic chip{\textendash}based optical frequency comb using soliton
  {Cherenkov} radiation.
\newblock {\em Science}, 351(6271):357--360.

\bibitem[\protect\astroncite{Brasch et~al.}{2019}]{brasch_nonlinear_2019}
Brasch, V., Obrzud, E., Lecomte, S., and Herr, T. (2019).
\newblock Nonlinear filtering of an optical pulse train using dissipative
  {Kerr} solitons.
\newblock {\em Optica}, 6(11):1386--1393.

\bibitem[\protect\astroncite{Coen et~al.}{1999}]{coen_convection_1999}
Coen, S., Tlidi, M., Emplit, P., and Haelterman, M. (1999).
\newblock Convection versus {Dispersion} in {Optical} {Bistability}.
\newblock {\em Phys. Rev. Lett.}, 83(12):2328--2331.

\bibitem[\protect\astroncite{Cole et~al.}{2018}]{cole_kerr-microresonator_2018}
Cole, D.~C., Stone, J.~R., Erkintalo, M., Yang, K.~Y., Yi, X., Vahala, K.~J.,
  and Papp, S.~B. (2018).
\newblock Kerr-microresonator solitons from a chirped background.
\newblock {\em Optica}, 5(10):1304--1310.

\bibitem[\protect\astroncite{Dong et~al.}{2020}]{dong_stretched-pulse_2020}
Dong, X., Yang, Q., Spiess, C., Bucklew, V.~G., and Renninger, W.~H. (2020).
\newblock Stretched-{Pulse} {Soliton} {Kerr} {Resonators}.
\newblock {\em Phys. Rev. Lett.}, 125(3):033902.

\bibitem[\protect\astroncite{Dutt et~al.}{2018}]{dutt_-chip_2018}
Dutt, A., Joshi, C., Ji, X., Cardenas, J., Okawachi, Y., Luke, K., Gaeta,
  A.~L., and Lipson, M. (2018).
\newblock On-chip dual-comb source for spectroscopy.
\newblock {\em Science Advances}, 4(3):e1701858.

\bibitem[\protect\astroncite{Firth and Scroggie}{1996}]{firth_optical_1996}
Firth, W.~J. and Scroggie, A.~J. (1996).
\newblock Optical {Bullet} {Holes}: {Robust} {Controllable} {Localized}
  {States} of a {Nonlinear} {Cavity}.
\newblock {\em Phys. Rev. Lett.}, 76(10):1623--1626.

\bibitem[\protect\astroncite{F{\"u}l{\"o}p
  et~al.}{2018}]{fulop_high-order_2018}
F{\"u}l{\"o}p, A., Mazur, M., Lorences-Riesgo, A., Helgason, {\'O}.~B., Wang,
  P.-H., Xuan, Y., Leaird, D.~E., Qi, M., Andrekson, P.~A., Weiner, A.~M., and
  Torres-Company, V. (2018).
\newblock High-order coherent communications using mode-locked dark-pulse
  {Kerr} combs from microresonators.
\newblock {\em Nature Commun.}, 9(1):1--8.

\bibitem[\protect\astroncite{Gohle et~al.}{2005}]{gohle_frequency_2005}
Gohle, C., Udem, T., Herrmann, M., Rauschenberger, J., Holzwarth, R.,
  Schuessler, H.~A., Krausz, F., and H{\"a}nsch, T.~W. (2005).
\newblock A frequency comb in the extreme ultraviolet.
\newblock {\em Nature}, 436(7048):234--237.

\bibitem[\protect\astroncite{Guo et~al.}{2017}]{guo_universal_2017}
Guo, H., Karpov, M., Lucas, E., Kordts, A., Pfeiffer, M. H.~P., Brasch, V.,
  Lihachev, G., Lobanov, V.~E., Gorodetsky, M.~L., and Kippenberg, T.~J.
  (2017).
\newblock Universal dynamics and deterministic switching of dissipative {Kerr}
  solitons in optical microresonators.
\newblock {\em Nature Phys.}, 13(1):94--102.

\bibitem[\protect\astroncite{Haelterman
  et~al.}{1992}]{haelterman_dissipative_1992}
Haelterman, M., Trillo, S., and Wabnitz, S. (1992).
\newblock Dissipative modulation instability in a nonlinear dispersive ring
  cavity.
\newblock {\em Opt. Commun.}, 91(5{\textendash}6):401--407.

\bibitem[\protect\astroncite{Helgason et~al.}{2020}]{helgason_dissipative_2020}
Helgason, {\'O}.~B., Arteaga-Sierra, F.~R., Ye, Z., Twayana, K., Andrekson,
  P.~A., Karlsson, M., Schr{\"o}der, J., and Torres-Company, V. (2020).
\newblock Dissipative {Kerr} solitons in photonic molecules.
\newblock {\em arXiv:2007.02608 [nlin, physics:physics]}.

\bibitem[\protect\astroncite{Hendry et~al.}{2018}]{hendry_spontaneous_2018}
Hendry, I., Chen, W., Wang, Y., Garbin, B., Javaloyes, J., Oppo, G.-L., Coen,
  S., Murdoch, S.~G., and Erkintalo, M. (2018).
\newblock Spontaneous symmetry breaking and trapping of temporal {Kerr} cavity
  solitons by pulsed or amplitude-modulated driving fields.
\newblock {\em Phys. Rev. A}, 97(5):053834.

\bibitem[\protect\astroncite{Hendry et~al.}{2019}]{hendry_impact_2019}
Hendry, I., Garbin, B., Murdoch, S.~G., Coen, S., and Erkintalo, M. (2019).
\newblock Impact of desynchronization and drift on soliton-based {Kerr}
  frequency combs in the presence of pulsed driving fields.
\newblock {\em Phys. Rev. A}, 100(2):023829.

\bibitem[\protect\astroncite{Herr et~al.}{2014}]{herr_temporal_2014}
Herr, T., Brasch, V., Jost, J.~D., Wang, C.~Y., Kondratiev, N.~M., Gorodetsky,
  M.~L., and Kippenberg, T.~J. (2014).
\newblock Temporal solitons in optical microresonators.
\newblock {\em Nature Photon.}, 8(2):145--152.

\bibitem[\protect\astroncite{Jang et~al.}{2015}]{jang_temporal_2015}
Jang, J.~K., Erkintalo, M., Coen, S., and Murdoch, S.~G. (2015).
\newblock Temporal tweezing of light through the trapping and manipulation of
  temporal cavity solitons.
\newblock {\em Nature Commun.}, 6(1):7370.

\bibitem[\protect\astroncite{Jang et~al.}{2013}]{jang_ultraweak_2013}
Jang, J.~K., Erkintalo, M., Murdoch, S.~G., and Coen, S. (2013).
\newblock Ultraweak long-range interactions of solitons observed over
  astronomical distances.
\newblock {\em Nature Photon.}, 7(8):657--663.

\bibitem[\protect\astroncite{Jang et~al.}{2016}]{jang_all-optical_2016}
Jang, J.~K., Erkintalo, M., Schr{\"o}der, J., Eggleton, B.~J., Murdoch, S.~G.,
  and Coen, S. (2016).
\newblock All-optical buffer based on temporal cavity solitons operating at 10
  {Gb}/s.
\newblock {\em Opt. Lett.}, 41(19):4526.

\bibitem[\protect\astroncite{Jang et~al.}{2018}]{jang_synchronization_2018}
Jang, J.~K., Klenner, A., Ji, X., Okawachi, Y., Lipson, M., and Gaeta, A.~L.
  (2018).
\newblock Synchronization of coupled optical microresonators.
\newblock {\em Nature Photon.}, 12(11):688--693.

\bibitem[\protect\astroncite{Joshi et~al.}{2016}]{joshi_thermally_2016}
Joshi, C., Jang, J.~K., Luke, K., Ji, X., Miller, S.~A., Klenner, A., Okawachi,
  Y., Lipson, M., and Gaeta, A.~L. (2016).
\newblock Thermally controlled comb generation and soliton modelocking in
  microresonators.
\newblock {\em Opt. Lett.}, 41(11):2565--2568.

\bibitem[\protect\astroncite{Karpov et~al.}{2019}]{karpov_dynamics_2019}
Karpov, M., Pfeiffer, M. H.~P., Guo, H., Weng, W., Liu, J., and Kippenberg,
  T.~J. (2019).
\newblock Dynamics of soliton crystals in optical microresonators.
\newblock {\em Nature Phys.}, 15(10):1071--1077.

\bibitem[\protect\astroncite{Kelley}{2003}]{kelley_solving_2003}
Kelley, C.~T. (2003).
\newblock {\em Solving {Nonlinear} {Equations} with {Newton}'s {Method}}.
\newblock Fundamentals of {Algorithms}. Society for Industrial and Applied
  Mathematics.

\bibitem[\protect\astroncite{Kim et~al.}{2019}]{kim_turn-key_2019}
Kim, B.~Y., Okawachi, Y., Jang, J.~K., Yu, M., Ji, X., Zhao, Y., Joshi, C.,
  Lipson, M., and Gaeta, A.~L. (2019).
\newblock Turn-key, high-efficiency {Kerr} comb source.
\newblock {\em Opt. Lett.}, 44(18):4475--4478.

\bibitem[\protect\astroncite{Kippenberg
  et~al.}{2018}]{kippenberg_dissipative_2018}
Kippenberg, T.~J., Gaeta, A.~L., Lipson, M., and Gorodetsky, M.~L. (2018).
\newblock Dissipative {Kerr} solitons in optical microresonators.
\newblock {\em Science}, 361(6402):eaan8083.

\bibitem[\protect\astroncite{Kobayashi
  et~al.}{1972}]{kobayashi_highrepetitionrate_1972}
Kobayashi, T., Sueta, T., Cho, Y., and Matsuo, Y. (1972).
\newblock High-repetition-rate optical pulse generator using a {Fabry}-{Perot}
  electro-optic modulator.
\newblock {\em Appl. Phys. Lett.}, 21(8):341--343.

\bibitem[\protect\astroncite{Kobayashi et~al.}{1988}]{kobayashi_optical_1988}
Kobayashi, T., Yao, H., Amano, K., Fukushima, Y., Morimoto, A., and Sueta, T.
  (1988).
\newblock Optical pulse compression using high-frequency electrooptic phase
  modulation.
\newblock {\em IEEE Journal of Quantum Electronics}, 24(2):382--387.

\bibitem[\protect\astroncite{Leo et~al.}{2010}]{leo_temporal_2010}
Leo, F., Coen, S., Kockaert, P., Gorza, S.-P., Emplit, P., and Haelterman, M.
  (2010).
\newblock Temporal cavity solitons in one-dimensional {Kerr} media as bits in
  an all-optical buffer.
\newblock {\em Nature Photon.}, 4(7):471--476.

\bibitem[\protect\astroncite{Leo et~al.}{2013}]{leo_dynamics_2013}
Leo, F., Gelens, L., Emplit, P., Haelterman, M., and Coen, S. (2013).
\newblock Dynamics of one-dimensional {Kerr} cavity solitons.
\newblock {\em Opt. Express}, 21(7):9180--9191.

\bibitem[\protect\astroncite{Li et~al.}{2020}]{li_experimental_2020}
Li, Z., Xu, Y., Coen, S., Murdoch, S.~G., and Erkintalo, M. (2020).
\newblock Experimental observations of bright dissipative cavity solitons and
  their collapsed snaking in a {Kerr} resonator with normal dispersion driving.
\newblock {\em Optica}, 7(9):1195--1203.

\bibitem[\protect\astroncite{Lilienfein
  et~al.}{2019}]{lilienfein_temporal_2019}
Lilienfein, N., Hofer, C., H{\"o}gner, M., Saule, T., Trubetskov, M., Pervak,
  V., Fill, E., Riek, C., Leitenstorfer, A., Limpert, J., Krausz, F., and
  Pupeza, I. (2019).
\newblock Temporal solitons in free-space femtosecond enhancement cavities.
\newblock {\em Nature Photon.}, 13(3):214--218.

\bibitem[\protect\astroncite{Lobanov et~al.}{2015}]{lobanov_frequency_2015}
Lobanov, V.~E., Lihachev, G., Kippenberg, T.~J., and Gorodetsky, M.~L. (2015).
\newblock Frequency combs and platicons in optical microresonators with normal
  {GVD}.
\newblock {\em Opt. Express}, 23(6):7713--7721.

\bibitem[\protect\astroncite{Lobanov et~al.}{2016}]{lobanov_harmonization_2016}
Lobanov, V.~E., Lihachev, G.~V., Pavlov, N.~G., Cherenkov, A.~V., Kippenberg,
  T.~J., and Gorodetsky, M.~L. (2016).
\newblock Harmonization of chaos into a soliton in {Kerr} frequency combs.
\newblock {\em Opt. Express}, 24(24):27382--27394.

\bibitem[\protect\astroncite{Lucas et~al.}{2018}]{lucas_spatial_2018}
Lucas, E., Lihachev, G., Bouchand, R., Pavlov, N.~G., Raja, A.~S., Karpov, M.,
  Gorodetsky, M.~L., and Kippenberg, T.~J. (2018).
\newblock Spatial multiplexing of soliton microcombs.
\newblock {\em Nautre Photon.}, 12(11):699--705.

\bibitem[\protect\astroncite{Lugiato and Lefever}{1987}]{lugiato_spatial_1987}
Lugiato, L.~A. and Lefever, R. (1987).
\newblock Spatial {Dissipative} {Structures} in {Passive} {Optical} {Systems}.
\newblock {\em Phys. Rev. Lett.}, 58(21):2209--2211.

\bibitem[\protect\astroncite{Luo et~al.}{2015}]{luo_spontaneous_2015}
Luo, K., Jang, J.~K., Coen, S., Murdoch, S.~G., and Erkintalo, M. (2015).
\newblock Spontaneous creation and annihilation of temporal cavity solitons in
  a coherently driven passive fiber resonator.
\newblock {\em Opt. Lett.}, 40(16):3735--3738.

\bibitem[\protect\astroncite{Maggipinto et~al.}{2000}]{maggipinto_cavity_2000}
Maggipinto, T., Brambilla, M., Harkness, G.~K., and Firth, W.~J. (2000).
\newblock Cavity solitons in semiconductor microresonators: {Existence},
  stability, and dynamical properties.
\newblock {\em Phys. Rev. E}, 62(6):8726--8739.

\bibitem[\protect\astroncite{Marin-Palomo
  et~al.}{2017}]{marin-palomo_microresonator-based_2017}
Marin-Palomo, P., Kemal, J.~N., Karpov, M., Kordts, A., Pfeifle, J., Pfeiffer,
  M. H.~P., Trocha, P., Wolf, S., Brasch, V., Anderson, M.~H., Rosenberger, R.,
  Vijayan, K., Freude, W., Kippenberg, T.~J., and Koos, C. (2017).
\newblock Microresonator-based solitons for massively parallel coherent optical
  communications.
\newblock {\em Nature}, 546(7657):274.

\bibitem[\protect\astroncite{Nielsen et~al.}{2019}]{nielsen_coexistence_2019}
Nielsen, A.~U., Garbin, B., Coen, S., Murdoch, S.~G., and Erkintalo, M. (2019).
\newblock Coexistence and interactions between nonlinear states with different
  polarizations in a monochromatically driven passive {Kerr} resonator.
\newblock {\em Phys. Rev. Lett.}, 123(1):013902.

\bibitem[\protect\astroncite{Obrzud et~al.}{2017}]{obrzud_temporal_2017}
Obrzud, E., Lecomte, S., and Herr, T. (2017).
\newblock Temporal solitons in microresonators driven by optical pulses.
\newblock {\em Nature Photon.}, 11(9):600.

\bibitem[\protect\astroncite{Obrzud et~al.}{2019}]{obrzud_microphotonic_2019}
Obrzud, E., Rainer, M., Harutyunyan, A., Anderson, M.~H., Liu, J., Geiselmann,
  M., Chazelas, B., Kundermann, S., Lecomte, S., Cecconi, M., Ghedina, A.,
  Molinari, E., Pepe, F., Wildi, F., Bouchy, F., Kippenberg, T.~J., and Herr,
  T. (2019).
\newblock A microphotonic astrocomb.
\newblock {\em Nature Photon.}, 13(1):31--35.

\bibitem[\protect\astroncite{Parra-Rivas
  et~al.}{2014}]{parra-rivas_effects_2014}
Parra-Rivas, P., Gomila, D., Mat{\'i}as, M.~A., Colet, P., and Gelens, L.
  (2014).
\newblock Effects of inhomogeneities and drift on the dynamics of temporal
  solitons in fiber cavities and microresonators.
\newblock {\em Opt. Express}, 22(25):30943--30954.

\bibitem[\protect\astroncite{Pasquazi
  et~al.}{2018}]{pasquazi_micro-combs:_2018}
Pasquazi, A., Peccianti, M., Razzari, L., Moss, D.~J., Coen, S., Erkintalo, M.,
  Chembo, Y.~K., Hansson, T., Wabnitz, S., Del{\textquoteright}Haye, P., Xue,
  X., Weiner, A.~M., and Morandotti, R. (2018).
\newblock Micro-combs: {A} novel generation of optical sources.
\newblock {\em Phys. Rep.}, 729:1--81.

\bibitem[\protect\astroncite{Pedaci et~al.}{2006}]{pedaci_positioning_2006}
Pedaci, F., Genevet, P., Barland, S., Giudici, M., and Tredicce, J.~R. (2006).
\newblock Positioning cavity solitons with a phase mask.
\newblock {\em Applied Physics Letters}, 89(22):221111.

\bibitem[\protect\astroncite{Riemensberger
  et~al.}{2020}]{riemensberger_massively_2020}
Riemensberger, J., Lukashchuk, A., Karpov, M., Weng, W., Lucas, E., Liu, J.,
  and Kippenberg, T.~J. (2020).
\newblock Massively parallel coherent laser ranging using a soliton microcomb.
\newblock {\em Nature}, 581(7807):164--170.

\bibitem[\protect\astroncite{Saule et~al.}{2019}]{saule_high-flux_2019}
Saule, T., Heinrich, S., Sch{\"o}tz, J., Lilienfein, N., H{\"o}gner, M.,
  deVries, O., Pl{\"o}tner, M., Weitenberg, J., Esser, D., Schulte, J.,
  Russbueldt, P., Limpert, J., Kling, M.~F., Kleineberg, U., and Pupeza, I.
  (2019).
\newblock High-flux ultrafast extreme-ultraviolet photoemission spectroscopy at
  18.4 {MHz} pulse repetition rate.
\newblock {\em Nature Commun.}, 10(1):458.

\bibitem[\protect\astroncite{Spencer
  et~al.}{2018}]{spencer_optical-frequency_2018}
Spencer, D.~T., Drake, T., Briles, T.~C., Stone, J., Sinclair, L.~C., Fredrick,
  C., Li, Q., Westly, D., Ilic, B.~R., Bluestone, A., Volet, N., Komljenovic,
  T., Chang, L., Lee, S.~H., Oh, D.~Y., Suh, M.-G., Yang, K.~Y., Pfeiffer, M.
  H.~P., Kippenberg, T.~J., Norberg, E., Theogarajan, L., Vahala, K., Newbury,
  N.~R., Srinivasan, K., Bowers, J.~E., Diddams, S.~A., and Papp, S.~B. (2018).
\newblock An optical-frequency synthesizer using integrated photonics.
\newblock {\em Nature}, 557(7703):81--85.

\bibitem[\protect\astroncite{Stern et~al.}{2020}]{stern_direct_2020}
Stern, L., Stone, J.~R., Kang, S., Cole, D.~C., Suh, M.-G., Fredrick, C.,
  Newman, Z., Vahala, K., Kitching, J., Diddams, S.~A., and Papp, S.~B. (2020).
\newblock Direct {Kerr} frequency comb atomic spectroscopy and stabilization.
\newblock {\em Science Advances}, 6(9):eaax6230.

\bibitem[\protect\astroncite{Suh and Vahala}{2018}]{suh_soliton_2018}
Suh, M.-G. and Vahala, K.~J. (2018).
\newblock Soliton microcomb range measurement.
\newblock {\em Science}, 359(6378):884--887.

\bibitem[\protect\astroncite{Suh et~al.}{2016}]{suh_microresonator_2016}
Suh, M.-G., Yang, Q.-F., Yang, K.~Y., Yi, X., and Vahala, K.~J. (2016).
\newblock Microresonator soliton dual-comb spectroscopy.
\newblock {\em Science}, 354(6312):600--603.

\bibitem[\protect\astroncite{Suh et~al.}{2019}]{suh_searching_2019}
Suh, M.-G., Yi, X., Lai, Y.-H., Leifer, S., Grudinin, I.~S., Vasisht, G.,
  Martin, E.~C., Fitzgerald, M.~P., Doppmann, G., Wang, J., Mawet, D., Papp,
  S.~B., Diddams, S.~A., Beichman, C., and Vahala, K. (2019).
\newblock Searching for exoplanets using a microresonator astrocomb.
\newblock {\em Nature Photon.}, 13(1):25--30.

\bibitem[\protect\astroncite{Taheri et~al.}{2015}]{taheri_soliton_2015}
Taheri, H., Eftekhar, A.~A., Wiesenfeld, K., and Adibi, A. (2015).
\newblock Soliton {Formation} in {Whispering}-{Gallery}-{Mode} {Resonators} via
  {Input} {Phase} {Modulation}.
\newblock {\em IEEE Photonics Journal}, 7(2):1--9.

\bibitem[\protect\astroncite{Trocha et~al.}{2018}]{trocha_ultrafast_2018}
Trocha, P., Karpov, M., Ganin, D., Pfeiffer, M. H.~P., Kordts, A., Wolf, S.,
  Krockenberger, J., Marin-Palomo, P., Weimann, C., Randel, S., Freude, W.,
  Kippenberg, T.~J., and Koos, C. (2018).
\newblock Ultrafast optical ranging using microresonator soliton frequency
  combs.
\newblock {\em Science}, 359(6378):887--891.

\bibitem[\protect\astroncite{Wabnitz}{1993}]{wabnitz_suppression_1993}
Wabnitz, S. (1993).
\newblock Suppression of interactions in a phase-locked soliton optical memory.
\newblock {\em Opt. Lett.}, 18(8):601--603.

\bibitem[\protect\astroncite{Wang}{2018}]{wang_temporal_2018}
Wang, Y. (2018).
\newblock {\em Temporal {Cavity} {Soliton} {Dynamics} in {Passive} {Kerr}
  {Resonators}}.
\newblock Doctoral Thesis, The University of Auckland.

\bibitem[\protect\astroncite{Wang et~al.}{2018a}]{wang_stimulated_2018}
Wang, Y., Anderson, M., Coen, S., Murdoch, S.~G., and Erkintalo, M. (2018a).
\newblock Stimulated {Raman} {Scattering} {Imposes} {Fundamental} {Limits} to
  the {Duration} and {Bandwidth} of {Temporal} {Cavity} {Solitons}.
\newblock {\em Phys. Rev. Lett.}, 120(5):053902.

\bibitem[\protect\astroncite{Wang et~al.}{2018b}]{wang_addressing_2018}
Wang, Y., Garbin, B., Leo, F., Coen, S., Erkintalo, M., and Murdoch, S.~G.
  (2018b).
\newblock Addressing temporal {Kerr} cavity solitons with a single pulse of
  intensity modulation.
\newblock {\em Opt. Lett.}, 43(13):3192--3195.

\bibitem[\protect\astroncite{Wang et~al.}{2017}]{wang_universal_2017}
Wang, Y., Leo, F., Fatome, J., Erkintalo, M., Murdoch, S.~G., and Coen, S.
  (2017).
\newblock Universal mechanism for the binding of temporal cavity solitons.
\newblock {\em Optica}, 4(8):855--863.

\bibitem[\protect\astroncite{Xu et~al.}{2020a}]{xu_spontaneous_2020}
Xu, G., Nielsen, A., Garbin, B., Hill, L., Oppo, G.-L., Fatome, J., Murdoch,
  S.~G., Coen, S., and Erkintalo, M. (2020a).
\newblock Spontaneous symmetry breaking of dissipative optical solitons in a
  two-component {Kerr} resonator.
\newblock {\em arXiv:2008.13776 [physics]}.

\bibitem[\protect\astroncite{Xu et~al.}{2020b}]{xu_harmonic_2020}
Xu, Y., Lin, Y., Nielsen, A., Hendry, I.,
  Coen, S.,  Erkintalo, M., Ma, H., Murdoch,
  S.~G., (2020b).
\newblock Harmonic and rational harmonic driving of microresonator soliton
  frequency combs.
\newblock {\em Optica}, 7(8):940--946.

\bibitem[\protect\astroncite{Xue et~al.}{2017}]{xue_microresonator_2017}
Xue, X., Wang, P.-H., Xuan, Y., Qi, M., and Weiner, A.~M. (2017).
\newblock Microresonator {Kerr} frequency combs with high conversion
  efficiency.
\newblock {\em Laser \& Photonics Reviews}, 11(1):1600276.

\bibitem[\protect\astroncite{Xue et~al.}{2015}]{xue_mode-locked_2015}
Xue, X., Xuan, Y., Liu, Y., Wang, P.-H., Chen, S., Wang, J., Leaird, D.~E., Qi,
  M., and Weiner, A.~M. (2015).
\newblock Mode-locked dark pulse {Kerr} combs in normal-dispersion
  microresonators.
\newblock {\em Nature Photon.}, 9(9):594--600.

\bibitem[\protect\astroncite{Xue et~al.}{2019}]{xue_super-efficient_2019}
Xue, X., Zheng, X., and Zhou, B. (2019).
\newblock Super-efficient temporal solitons in mutually coupled optical
  cavities.
\newblock {\em Nature Photon.}, 13(9):616--622.

\bibitem[\protect\astroncite{Yi et~al.}{2015}]{yi_soliton_2015}
Yi, X., Yang, Q.-F., Yang, K.~Y., Suh, M.-G., and Vahala, K. (2015).
\newblock Soliton frequency comb at microwave rates in a high-{Q} silica
  microresonator.
\newblock {\em Optica}, 2(12):1078.

\end{thebibliography}

\end{document}